\documentclass[lettersize,journal]{IEEEtran}
\usepackage{amsmath,amsfonts}
\usepackage{algorithmic}
\usepackage{algorithm}
\usepackage{array}
\usepackage{textcomp}
\usepackage{stfloats}
\usepackage{url}
\usepackage{verbatim}
\usepackage{graphicx}
\usepackage{subfigure}
\usepackage{cite}
\usepackage{multirow}
\usepackage{multicol}
\usepackage[switch]{lineno}
\usepackage{color}

\usepackage[colorlinks,linkcolor=blue]{hyperref}

\begin{document}

\title{Crucial Feature Capture and Discrimination for Limited Training Data SAR ATR}

\author{Chenwei Wang,~\IEEEmembership{Student Member,~IEEE,}
        Siyi Luo,
        Jifang Pei,~\IEEEmembership{Member,~IEEE,}\\
        Yulin Huang,~\IEEEmembership{Senior Member,~IEEE,}
        Yin Zhang,~\IEEEmembership{Member,~IEEE,}
        and Jianyu Yang,~\IEEEmembership{Member,~IEEE}

\thanks{This work was supported by the National Natural Science Foundation of China under Grants 61901091 and 61901090.}
\thanks{\emph{Corresponding author: Jifang Pei.}}
\thanks{The authors are with the Department of Electrical Engineering, University of Electronic Science and Technology of China, Chengdu 611731, China (e-mail: peijfstudy@126.com; dbw181101@163.com).}}

\markboth{Journal of \LaTeX\ Class Files,~Vol.~14, No.~8, August~2021}%
{Shell \MakeLowercase{\textit{et al.}}: A Sample Article Using IEEEtran.cls for IEEE Journals}

\maketitle

\begin{abstract}
Although deep learning-based methods have achieved excellent performance on SAR ATR, the fact that it is difficult to acquire and label a lot of SAR images makes these methods, which originally performed well, perform weakly. This may be because most of them consider the whole target images as input, but the researches find that, under limited training data, the deep learning model can't capture discriminative image regions in the whole images, rather focus on more useless even harmful image regions for recognition. Therefore, the results are not satisfactory.
In this paper, we design a SAR ATR framework under limited training samples, which mainly consists of two branches and two modules, global assisted branch and local enhanced branch, feature capture module and feature discrimination module. 
In every training process, the global assisted branch first finishes the initial recognition based on the whole image. Based on the initial recognition results, the feature capture module automatically searches and locks the crucial image regions for correct recognition, which we named as the golden key of image. 
Then the local extract the local features from the captured crucial image regions.
The feature discrimination module provide a hybrid loss to enhance the intra-class compactness and inter-class separability of the overall features and local features and solve the poor performance of the global assisted branch after the initialization. 
Finally, the overall features and local features are input into the classifier and dynamically weighted using the learnable voting parameters to collaboratively complete the final recognition under limited training samples.
The model soundness experiments demonstrate the effectiveness of our method through the improvement of feature distribution and recognition probability. 
The experimental results and comparisons on MSTAR and OPENSAR show that our method has achieved superior recognition performance. We will release our code and more experimental results at  \url{https://github.com/cwwangSARATR/SARATR_FeaCapture_Discrimination}.
\end{abstract}


\section{Introduction}

\IEEEPARstart{S}{ynthetic} aperture radar (SAR) is an active remote sensor, which can provide high-resolution, day-and-night, and weather-independent images for a multitude of civilian and military applications \cite{intro1}. Automatic target recognition (ATR) applies computer processing capabilities to predict the class of an unknown target, which has become a very challenging problem in SAR application field \cite{intro2}. A standard architecture for efficient SAR ATR consists of three stages: detection, discrimination, and classification. Each stage is inclined to perform more complicated and elaborate processing than the prior stage and to select candidate objects for the next stage of processing. 
There have been many outstanding methods for SAR ATR, which can be divided into template-based and model-based approaches \cite{reff1, reff2, reff3, my1}.

Due to deep learning as the fast-growing of the big data era trend, it is proving to be a remarkably successful tool, sometimes even surpassing humans in solving highly computational tasks \cite{ATR3}, \cite{reff4}. In recent years, researchers all around the world have made increasing progress in solving the bottleneck problem of SAR ATR with the help of deep learning-based algorithms \cite{reff5, wang2023entropy,wang2022recognition,wang2023sar,wang2022global,wang2022semi,reff6}. For example, Chen and Wang \cite{reff7} introduced CNNs into SAR ATR and tested them on the standard ATR moving and stationary target acquisition and recognition (MSTAR) data set. Wagner \cite{reff8} suggested using a CNN to first extract feature vectors and then feed them to an SVM for classification. Chen et al. \cite{ATR1} proposed an all-convolutional network consisting of sparsely connected layers without using fully connected layers to avoid overfitting. 

However, deep learning networks require large amounts of labeled training data. Therefore, an intrinsic problem of recognition performance degradation emerges when the number of available labeled SAR target images is limited. This will lead to serious overfitting and, consequently, significantly limit or destroy the performance of the deep learning models.
Meanwhile, the problem of insufficient SAR images is basic and crucial in SAR application. For example, for earthquake rescue or debris flow observation, SAR can provide special imaging capability to help the rescue. However, there are various buildings or vehicles in the scene, while different shapes or sizes of these things will finally lead to the problem of insufficient SAR images, let alone the various platforms and imaging conditions.

\begin{figure*}[htb]
\centering
\includegraphics[width=1\textwidth]{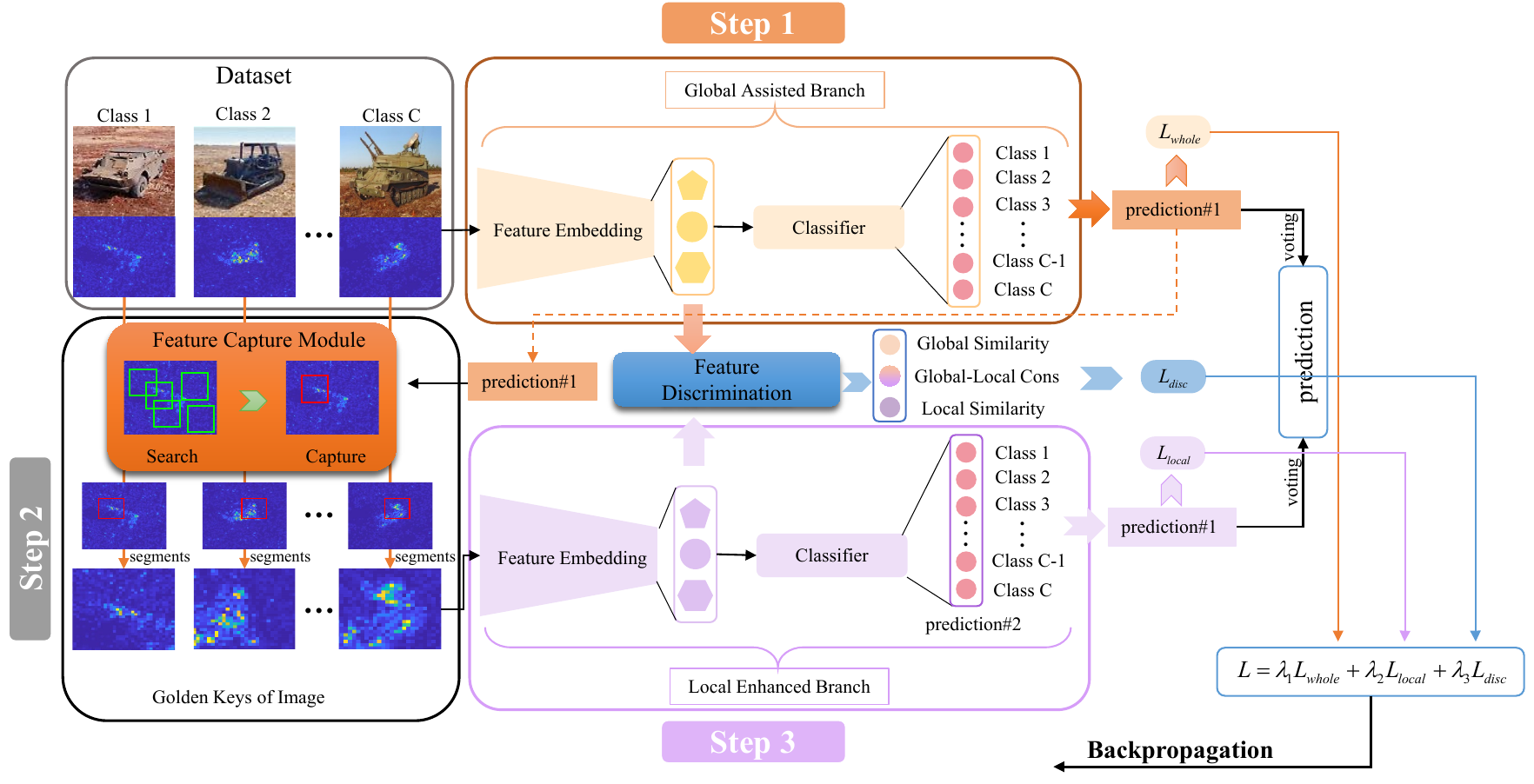}
\caption{Framework of SCDR. Left part is the whole framework of SCDR. Upper branch of left part perceive the whole SAR image and the bottom branch of left part finishes the recognition only based on the crucial features extracted by the feature capture module.
Right part is the function of the discrimination for the features. Yellow pentagons and circles are from the upper branch, lavender pentagons and circles are from the bottom branch. Pentagons and circles are features from different class types.}
\label{fig2}
\end{figure*}

Currently, Most methods focus on improving recognition performance by training the network, provided that available data is sufficient or even adequate. While for the condition with severe lack of data, which is called the limited training sample condition, the existing methods usually input the whole target image as a whole to train a recognition network \cite{add1, add2, add3}. Generally, a recognition network consists of a feature extractor and a classifier. The features corresponding to the whole image are extracted first, and then imported as a whole into the classifier to finish recognition\cite{fslmodel1,wang2020deep,wang2022sar,wang2021multiview,wang2019parking,wang2021deep,fslmodel2,fslmodel3}.

However, it has been found that, under limited training data, the deep learning model can't capture the crucial image part in the whole image, rather focus on the useless or even harmful image part for recognition \cite{deeplearning,attention1,attention2,wang2020multi,li2023panoptic,liang2023efficient,attention3,zhang2020deepemd,vinyals2016matching}. The label information is limited so that the model can't obtain enough information to decide where to look at in the whole SAR images and extract the discriminative feature, fail to achieve precise recognition performance. 
Thus, under the limited training SAR samples, the target recognition method based on the whole image is not ideal since the extracted overall features contain a large number of useless or even harmful local-image-region features rather than the discriminative feature from the crucial image region.

Since only a part of the image region is relatively discriminative and contributes mainly to the recognition performance, but under limited training data, the model can't capture this image part. Thus, we may assume that it is possible to construct a network structure that can automatically search and capture the discriminative image region from the whole image, focus on extracting features from this region, and discard other local image regions automatically. In this way, the model can capture the crucial image part for recognition and improve the effectiveness and discrimination of the overall feature of the whole image to achieve accurate recognition performance under limited training data.
We names such discriminative image region as the golden key of image under limited training data SAR ATR.
So as we can obtain a network that is more suitable for recognition under limited training SAR sample conditions, improve generalization ability and better enhance recognition performance.

Therefore, we propose a SAR ATR method to search and capture discriminative image region (SCDR), the method we designed can be illustrated as follows, as shown in Fig. \ref{fig2}.

1) First, it inputs the whole image to the global assisted branch, extracts the overall features and inputs them into the classifier to complete the recognition, as shown in the yellow branch in Fig. \ref{fig2}.

2) The proposed feature capture module is then used to obtain the weights assigned to different spatial location features when the global assisted branch correctly recognizes, and normalize them to the range of 0-1. The spatial location features with high weight is the crucial image region for correct recognition, which is the so-called golden key of image.
With the mean value of all weights being the threshold, the golden key is binarized to form a mask, which is then element-wise multiplied by the whole image to cut out the golden key as shown in the orange square in Fig. \ref{fig2}.

3) For the segmented golden key, the depth features are extracted individually in the local enhanced branch, as shown in the purple branch in Fig. \ref{fig2}.

4) As shown in the blue squares in Fig. \ref{fig2}, the intra-class compactness and inter-class separateness of the overall and local features of the yellow and purple branches are enhanced in feature discrimination module. Furthermore, the feature discrimination module can help the model capture the crucial image part when the model performs poorly at the start of training.

Finally, the yellow and purple branches extract the features of the whole image and the golden key respectively. The features are then input into the classifier to get the probability distribution of recognition respectively. The learnable parameters are used to voting weight the two probability distributions dynamically to get the final recognition probability distribution. Therefore, the yellow and purple branches can collaborate to complete the final recognition.

In this way, our SCDR can automatically search and capture the discriminative local image region in the whole image, discard other useless or even harmful image regions in the whole image, and focus on extracting features for this discriminative image region to complete the final recognition in cooperation with the overall features, which improves the recognition performance of the overall method and solves the difficulty of target recognition under limited training SAR samples in a targeted manner. 

The working mechanism of the SCDR is similar to the mechanism of the human brain to recognize objects in the real world. The SCDR first identifies the class type of SAR image by observing the whole SAR image and finds the crucial structures or features of the target in the image. Then the SCDR further focuses on and discriminates the found crucial features, and finally comprehensively considers the whole images and crucial features of the target to achieve the recognition of the target classes. 

The main contributions of this paper are summarized as follows:

(1) A novel framework for limited training data which comprehensively considers the whole image and crucial features to recognize the target classes is proposed. It not only perceives the whole target images, but also capture the crucial features and enhances the effectiveness of these crucial features for recognition.

(2) A feature capture module is proposed to locate the crucial features, and a feature discrimination module is proposed to make the overall and local features with more compactness and inter-class separability and solve the bad initialization.

(3) The SCDR achieves the state-of-the-art performance of recognition on MSTAR and OpenSARship data sets with limited training data. The method soudness verification also validate the effectiveness of our methods. 
 
The remainder of this paper is organized as follows. The related works are introduced in Section \uppercase\expandafter{\romannumeral2}. The details of the SCDR is presented in Section \uppercase\expandafter{\romannumeral3}.  The effectiveness of the proposed method are validated by experiments in Section \uppercase\expandafter{\romannumeral4}, and Section \uppercase\expandafter{\romannumeral5} gives a conclusion.

\section{Related Works}
\begin{figure}
\centering
\includegraphics[width=0.40\textwidth]{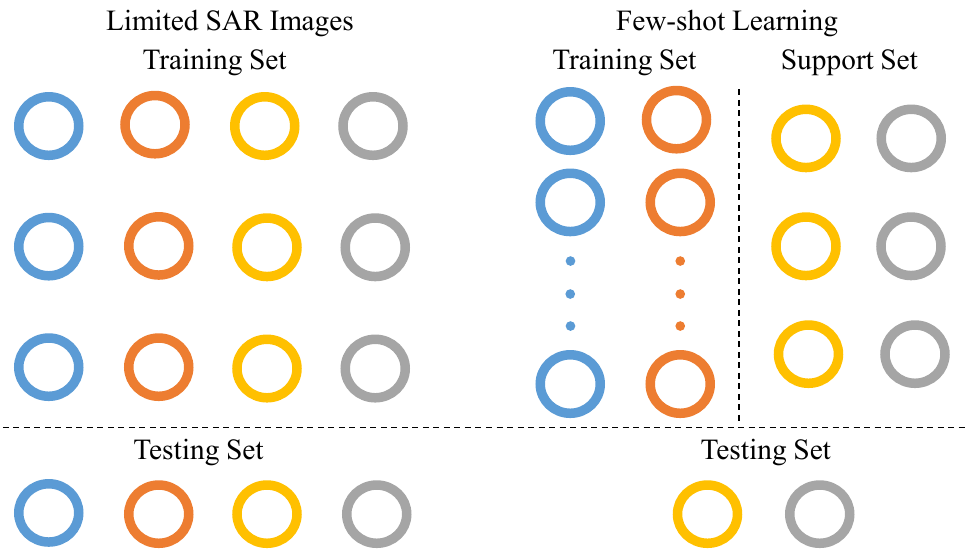}
\caption{Limited SAR images and FSL. One example of 3 images for each class for limited SAR images and FSL, the circles of the different colors denote the images of the different labeled target classes.}
\label{relatedwork}
\end{figure}

Many researches of the insufficient SAR images in SAR ATR have been carried out in recent years. These researches can be divided into two types based on the numbers of the limited training samples and problem formulation, the problem of the limited training data and the problem of FSL. In this section, the researches of the limited training data and FSL in SAR ATR are introduced as follows.
Besides, the training and testing settings of the limited training data and FSL are as shown in Fig. \ref{relatedwork}.

Limited training sample problem: In the problem of limited SAR images, there are one training dataset and one testing dataset. The sample number of every target class type in the training dataset is limited. For example, if there are 10 SAR images for all 10 class types, the training dataset has $10 \times 10 = 100$ SAR images in total. The models just employ 100 SAR images to train and test. Sometimes, some methods also construct a semi-supervised structure to employ the resting images as unlabeled samples besides the $K$ labeled images in training. Under this condition, the input information to model is more than the $K$ labeled images.

For example, Wang et al. \cite{reduce1} designed a semi-supervised learning framework mainly containing a self-consistent augmentation rule to utilize unlabeled data during training for the limited SAR images. Zhang et al. \cite{fsldata4} adopted the feature augmentation and ensemble learning strategies to concatenate cascaded features from optimally selected convolutional layers to obtain more comprehensive representation information from limited data. Sun et al. \cite{fsldata3} introduced an attribute-guided transfer learning method employing an angular rotation generative network, where the shared attribute between the source and target domains is the target aspect angle, to address the problem of the lack of training data at different aspect angles. Zhang et al. \cite{readd1} designed a semi-supervised transfer learning method for limited SAR training data, which adopts learned parameters from a pre-trained GAN, achieving up to 23.58\% accuracy improvement compared with other random-initialized models. Besides, they also proposed another transfer learning method for limited SAR training data, where the pre-trained layers are reused to transfer the generic knowledge \cite{readd2}.

FSL problem: In the problem of FSL, there is one support dataset with a limited number of samples for each class type, which plays the same role as the training dataset in limited SAR images. There is also one training dataset with a sufficient number of samples for other class types. The class types are disjoint between the training dataset and support dataset. Take the MSTAR dataset as an example, total of 10 class types, for 5-way 10-shot of FSL, 5-way means 5 class types in the support dataset, 10-shot means 10 SAR images for each type of 5 class types, there are $5 \times 10 = 50$ SAR images in the support dataset. Meanwhile, there is a training dataset that has sufficient SAR images of the remaining 5 class types disjoint from the support dataset.

Fu et al. \cite{fslmodel1} adopted three transfer learning methods and meta-learning in the few-shot SAR target classification to mitigate the difficulties with few training data. 
Wang et al. \cite{fslmodel2} proposed a hybrid inference network (HIN) containing an embedding network as the first stage and a hybrid inference strategy as the second stage, which obtained good results in the case of three-target classification for FSL. 
They also proposed a convolutional bidirectional long short-term memory prototypical network that was trained to map SAR images into a new feature space and utilized Euclidean distance to obtain the recognition results, achieving over 90\% recognition accuracy in the 5-shot scenario \cite{readd3}. 
\cite{fslmodel3} then proposed an FSL approach that uses a convolutional bidirectional long short-term memory network to extract azimuth-insensitive features and further improved the performance. 
Yang et al. \cite{fslmodel4} proposed a mixed loss graph attention network (MGA-NET) containing a data augmentation module, embedding network, and multilayer graph attention network, which increased the classification accuracy in 3-way 1-shot and 3-way 5-shot scenarios. Rostami et al. \cite{readd4} introduced a deep transfer learning method for FSL by transferring knowledge from the optical domain through learning a shared invariant cross-domain embedding space for discrimination. Wang et al. \cite{readd5} proposed an attribute-guided multi-scale prototypical network (AG-MsPN) combined with sub-band decomposition, obtaining superior performance in few-shot case.

Although these related researches have improved the recognition performance of SAR ATR under limited training data and FSL, some of them regard the whole SAR images as the input, the useless or even harmful image region inevitably limited their performance.
In this paper, our method focuses on the problem of the limited training data, it searches and captures the crucial image region in the whole image, which is named as the golden key of image under limited training data.
Then our method focuses on extracting features from the golden key of image to complete the final recognition in cooperation with the overall features. In this way, it can tackle with the core problem of limited training samples in SAR ATR in a targeted manner.

\section{Proposed Method}
In this section, the proposed framework of the SCDR is described in detail and the feature capture module and the feature discrimination module are also presented.

\subsection{Framework of SCDR}
To help the network search and capture crucial features and improve the discrimination of these features, first, we should know which local regions of the image are important for the recognition of the network. So as these image region can be regarded as the golden key of image.
Then, the discrimination or effectiveness of these features should be improved by enhancing the compactness of the features in the same class and the separability of the features among different classes. Finally, the core problem of limited training samples in SAR ATR can be solved.

As shown in Fig. \ref{fig2}, the whole framework of SCDR is constructed by two branches and one core module: the upper yellow branch with feature capture module, the bottom lavender branch to recognize from the local features, the blue discrimination module to enhance the feature effectiveness.
The upper yellow branch with feature capture module is designed for searching the crucial features of the images and capturing these features which are important for the recognition.
The bottom lavender branch is to judge whether the right recognition can be obtained only by these features to measure the effectiveness of these features and the capability of the upper branch.
The blue discrimination module can calculate the cosine similarity between the feature maps of the two branches and enhance the inner-class compactness and inter-class separability of the features maps of the two branches, which can finally improve the discrimination of these features.

By inputting the whole SAR image into the upper branch with feature capture module, the recognition results on whole images are obtained to calculate one loss of recognition ${{L}_{whole}}$, and the corresponding crucial features of the input images are output as the input of the bottom branch. Then the bottom branch also outputs the recognition results on these crucial features to calculate another loss of recognition ${{L}_{local}}$, and the feature maps of the two branches are input into the discrimination module to calculate the loss of the feature discrimination ${{L}_{disc}}$. The final loss can be calculated as follows:
\begin{align}
{L={{\lambda }_{1}}{{L}_{whole}}+{{\lambda }_{2}}{{L}_{local}}+{{\lambda }_{3}}{{L}_{disc}}}
\end{align}
where ${\lambda_1}$, ${\lambda_2}$ and ${\lambda_3}$ are the weighting coefficients.

As shown in Fig. \ref{fig3}, the process of the SCDR is also described in detail. Given a SAR image ${\mathbf{x}}_{i}$ of $y_i$ class type, by inputting ${\mathbf{x}}_{i}$ into the upper branch, the output feature maps of the feature embedding in the upper branch is $\mathbf{M}_{i}^{whole}\left( {\mathbf{x}}_{i} \right)\in {{\mathbb{R}}^{h\times w\times c}}$, and the classifier gives the class predictions ${{p}_{whole}}\left( {{y}_{j}}|{{\mathbf{x}}_{i}} \right)$, the probability of the sample ${\mathbf{x}}_{i}$ classified to $j\text{th}$ class. The loss of recognition ${{L}_{whole}}$ can be obtained as 
\begin{align}
{{{L}_{whole}}\left( {{\mathbf{x}}_{i}} \right) = -\sum\limits_{j=1}^{K}{{{y}_{j}}}\text{log}\left( {{p}_{whole}}\left( {{y}_{j}}|{{\mathbf{x}}_{i}} \right) \right)}
\end{align}
where $K$ is the class number.

Then, the feature capture module employs the predictions ${{{p}_{whole}}\left( {{y}_{j}}|{{\mathbf{x}}_{i}} \right)}$ and the feature maps ${{\bf{M}}_i^{whole}\left( {{\mathbf{x}}_{i}} \right)}$ to capture the crucial features in the input SAR images, ${{\mathbf{x}}_{i}^{local}}$, which is important for the recognition of the upper branch.

These features ${{\mathbf{x}}_{i}}^{local}$ in the input SAR images are severed as the input of the bottom branch. Then the feature maps $\mathbf{M}_{i}^{local}\left( {{\mathbf{x}}_{i}} \right)\in {{\mathbb{R}}^{h\times w\times c}}$ and the class predictions ${{p}_{local}}\left( {{y}_{j}}|{{\mathbf{x}}_{i}} \right)$ are also output by the feature embedding and the classifier in the bottom branch. The loss of recognition ${{L}_{local}}$ can be obtained as follows:
\begin{align}
{{{L}_{local}}\left( {{\mathbf{x}}_{i}} \right) = -\sum\limits_{j=1}^{K}{{{y}_{j}}}\text{log}\left( {{p}_{local}}\left( {{y}_{j}}|{{\mathbf{x}}_{i}} \right) \right)}
\end{align}

Through the process and two losses above, the upper and bottom branches can recognize the target classes on the whole images and crucial features. The performance of the bottom branch can be treated as the distinguishing level of these crucial features.
The upper branch is optimized by the recognition errors by the whole images in the upper branch and the crucial features in the bottom branch.
Through the optimization in the training, our method can enhance the effectiveness of these crucial features and improve the recognition performance.

\subsection{Feature Capture Module}

The feature capture module aims to search and capture crucial features in images for the final recognition. When inputting a whole SAR image to the deep learning models, the deep learning models classify the images into a specific target class based on the extracted feature maps. These feature maps are from different spatial local regions of the images and play different roles in the recognition. The feature capture module relies on the final recognition results to find crucial image regions for the correct prediction and the corresponding weights for crucial features. This is the reasons that we named these image regions as the golden key of image under limited training data. Motivated by \cite{CAM}, the process of the feature capture module is designed as two stages as follows.

The first stage is to search crucial features for the final recognition in the upper branch. Through the final class prediction, the feature maps can be weighted integrated together to find crucial features. The process of the weighted integration is described as follow.
Given an input SAR image ${{\mathbf{x}}_{i}}$ of $y_i$ class type and the feature maps before the classifier, $\mathbf{M}\left( {{\mathbf{x}}_{i}} \right)\in {{\mathbb{R}}^{h\times w\times c}}$, the feature maps ${\bf{M}}\left( {{\mathbf{x}}_{i}} \right)$ firstly go through the global average pooling to integrate global features into $\mathbf{F}\left( {{\mathbf{x}}_{i}} \right)\in {{\mathbb{R}}^{1\times 1\times c}}$, and the output of the classifier go through one SoftMax to get the final prediction $\mathbf{pred}\in {{\mathbb{R}}^{1\times K}}$ on the class of ${{\mathbf{x}}_{i}}$.
Then the prediction $\mathbf{pred}\in {{\mathbb{R}}^{1\times K}}$ is employed to choose the corresponding weights ${{\mathbf{W}}_{FC}}$ of the full connection layer as
\begin{align}
{\mathbf{W}_{FC}^{p}={{\mathbf{W}}_{FC}}\left[ \text{argmax}\left( \mathbf{pred} \right) \right] = \left\{ {{w}_{1}},{{w}_{2}},\ldots {{w}_{c}} \right\}}
\end{align}
where $\left[ {{\rm{argmax}}\left( {{ \cdot}} \right)} \right]$ means the corresponding index of maximum value, $K$ means the class number of the targets, ${{w}_{1}},{{w}_{2}},\ldots {{w}_{c}}$ means the parameters from the full connected layers.

The second stage is to capture the corresponding local parts in the input image based on the crucial features from the first stage, and cut out the local parts from the input image to input into the bottom branch. 
Then crucial features to be captured, ${{\mathbf{M}}_{fea_capture}}\in {{\mathbb{R}}^{h\times w}}$, can be calculated by
\begin{align}
{{{\mathbf{M}}_{fea_capture}} = \operatorname{reshape}\left( \sum\limits_{j=1}^{c}{{{w}_{j}}{{\mathbf{M}}_{j}}\left( {{\mathbf{x}}_{i}} \right)} \right)}
\end{align}
where ${\mathop{\rm reshape}\nolimits} \left(  \cdot  \right)$ means reshaping these local features into the size of the input image ${\mathbf{x}}_{i}$. The local parts to be captured, ${\mathbf{I}_{local}}$, are computed by
\begin{align}
\label{Ilocal}
{{\mathbf{I}_{local}} = \mu B\left( {{\mathbf{M}}_{fea_capture}},\varphi  \right) \odot  {\mathbf{x}}_{i}}
\end{align}
where ${\mathbf{I}_{local}}$ means the input for the bottom branch, $B\left( {{\cdot}} \right)$ means binarization with threshold $ \varphi $, and $\mu $ means the weight for the feature capture module, $ \odot$ means Hadamard product.

Through the feature capture module, the bottom branches can force the upper branch to locate the golden keys of image. For further enhancement of crucial features, the discrimination module is proposed and described in detail as follows.

\subsection{Feature Discrimination Module}

For the further enhancement of the crucial features in inner-class compactness and inter-class separability, feature discrimination module is necessary for the limited training sample in SAR ATR.

As shown in Fig. \ref{fig2}, the feature discrimination module aims to force the direction of the feature vectors of the same class closer and the direction of the feature vectors of the different classes more distant. To achieve this goal, the discrimination module calculates the cosine similarities of the features and organizes a novel loss with a margin of the feature directions between the different classes and the same class.

The discrimination module has two stages as shown in Fig. \ref{fig3}: find the hardest negative image and the hardest positive image, calculate the discrimination loss. In the first stage, for an input SAR image ${\mathbf{x}}_{i}$ of $y_i$ class type and the feature maps of the branches, $\mathbf{M}_{i}^{whole}\left( {\mathbf{x}}_{i} \right)$ and $\mathbf{M}_{i}^{local}\left( {\mathbf{x}}_{i} \right)$, in all inputted images $\left\{ {{\mathbf{x}}_{1}^{1}, {\mathbf{x}}_{1}^{2}, \ldots , {\mathbf{x}}_{K}^{N}} \right\} \in {\mathbb{R}^{K \times N}}$, $K$ classes and ${N}$ samples each class, the discrimination module firstly calculates the cosine similarity of all the features. The cosine similarity is computed as

\begin{align}
\begin{split}
{sim_{\cos }}\left( {{{\mathbf{x}}_{{i}}}{{,}}{{\mathbf{x}}_{{j}}}} \right){{ = }}\frac{{f\left( {{\mathbf{M}}_i^{whole}\left( {{{\mathbf{x}}_{{i}}}} \right)} \right)}}{{{{\left\| {f\left( {{\mathbf{M}}_i^{whole}\left( {{{\mathbf{x}}_{{i}}}} \right)} \right)} \right\|}_2}}} \cdot \frac{{f\left( {{\mathbf{M}}_j^{local}\left( {{{\mathbf{x}}_{{j}}}} \right)} \right)}}{{{{\left\| {f\left( {{\mathbf{M}}_j^{local}\left( {{{\mathbf{x}}_{{j}}}} \right)} \right)} \right\|}_2}}}
\end{split}
\end{align}
where $ {\left\|  \cdot  \right\|_2} $ means the $\rm{L}_2$-norm and $ f\left(  {\cdot}  \right) $ means flattening feature maps.

Then the rank list of the cosine similarity is obtained as $\mathbf{Score}$, the discrimination module tries to find the most similar features among all the local features of different classes named hardest negative ${{\mathbf{M}}_{neg\_hard}^{local}\left( {{{\mathbf{x}}_{i}}} \right)}$, and the least similar features of the same class named hardest positive ${{\mathbf{M}}_{pos\_hard}^{local}\left( {{{\mathbf{x}}_{i}}} \right)}$.
\begin{align}
{{{\mathbf{M}}_{neg\_hard}^{local}\left( {{{\mathbf{x}}_{{i}}}} \right)}\leftarrow sort\left( \mathbf{Score},descending \right)[0]}
\end{align}
\begin{align}
{{{\mathbf{M}}_{pos\_hard}^{local}\left( {{{\mathbf{x}}_{{i}}}} \right)}\leftarrow sort\left( \mathbf{Score},ascending \right)[0]}
\end{align}
where $sort\left(  \cdot  \right)$ means sort by descending or ascending. The hardest negative is the first features after sorting by descending and the hardest positive is the first features after sorting by ascending.

The second stage is to calculate the discrimination loss as follows:
\begin{align}
{{{L}_{disc}} = \max\left({{l}_{neg}}+ \psi -{{l}_{pos}},0 \right)}
\end{align}
where ${{l_{neg}}}$ means the similarity between ${{{\mathbf{x}}_{i}}}$ and the most similar samples of different classes $\mathbf{M}_{neg\_hard}^{local}$, ${{l_{pos}}}$ means the similarity between ${{{\mathbf{x}}_{{i}}}}$ and the least similar samples of different classes $\mathbf{M}_{pos\_hard}^{local}$, the constant $\psi$ means the margin of the similarity between ${{l_{neg}}}$ and ${{l_{pos}}}$. In this way, the discrimination loss can achieve that the similarity of the features from the same classes are larger $\psi$ than the similarity of the features from the different classes.

${{l_{neg}}}$ and ${{l_{pos}}}$ are calculated as
\begin{align}
{l_{neg}}{\text{ = }}\sum {\frac{{f\left( {{\mathbf{M}}_i^{whole}\left( {{x_i}} \right)} \right)}}{{{{\left\| {f\left( {{\mathbf{M}}_i^{whole}\left( {{x_i}} \right)} \right)} \right\|}_2}}} \odot \frac{{f\left( {{\mathbf{M}}_{neg\_hard}^{local}} \right)}}{{{{\left\| {f\left( {{\mathbf{M}}_{neg\_hard}^{local}} \right)} \right\|}_2}}}} 
\end{align}
\begin{align}
{l_{pos}}{\text{ = }}\sum {\frac{{f\left( {{\mathbf{M}}_i^{whole}\left( {{x_i}} \right)} \right)}}{{{{\left\| {f\left( {{\mathbf{M}}_i^{whole}\left( {{x_i}} \right)} \right)} \right\|}_2}}} \odot \frac{{f\left( {{\mathbf{M}}_{pos\_hard}^{local}} \right)}}{{{{\left\| {f\left( {{\mathbf{M}}_{pos\_hard}^{local}} \right)} \right\|}_2}}}} 
\end{align}
where $\sum\cdot$ means one function that sums all vector elements.

The role of the discrimination loss can be treated as finding the most discriminable local parts among the different classes and the most compact or common local features in the same class.

The proposed SCDR firstly utilized the two branches to recognize the target classes on the whole images and the local parts of images separately. The feature capture module is proposed to find the crucial local parts.
Then the discrimination module is proposed to enhance the inner-class compactness and inter-class separability of the local features.
Through the above innovations, our method can achieve superior recognition performance under the condition of limited training SAR data.

\section{Experiments and Results}

In this section, we validate the effectiveness and robustness of the proposed method on the MSTAR and OpenSARShip datasets. First, the two datasets are employed to evaluate the method and the corresponding preprocess are introduced in detail. 
Then, the testing sample heatmaps of the network without/with our method are shown to validate the method soundness. The heatmaps represent the weight of each spatial location on the testing image for the correct recognition. 
In our method, the local image regions are generated by binarizing the heatmaps.
At the same time, the experimental results under the standard operating condition (SOC) and extended operating conditions (EOCs) are demonstrated. The recognition results on the OpenSARShip dataset are also given. Finally, the comparisons with other SAR ATR methods are presented.

\subsection{Dataset and Network Configuration}

\begin{figure}
\centering
\includegraphics[width=0.48\textwidth]{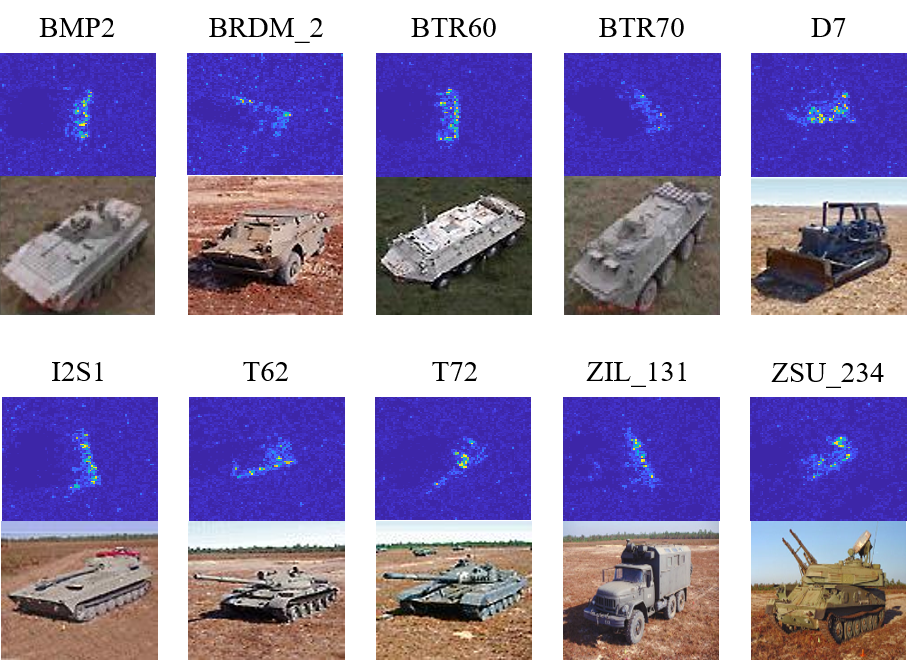}
\caption{SAR images and corresponding optical images of targets.}
\label{sampleMSTAR}
\end{figure}

\begin{figure}
\centering
\includegraphics[width=0.42\textwidth]{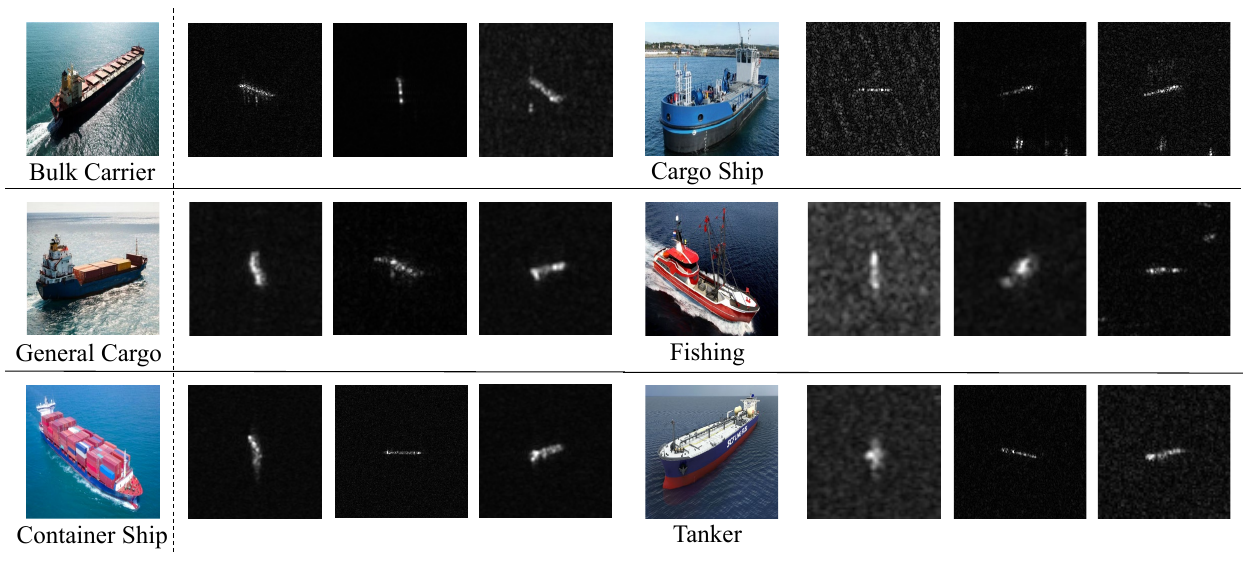}
\caption{SAR images and corresponding optical images of three-class targets in the OpenSARShip dataset.}
\label{sampleOPEN}
\end{figure}

In the experiment, we choose two benchmark datasets for SAR ATR, the moving and stationary target acquisition and recognition (MSTAR) and the OpenSARShip, to evaluate our proposed SCDR.

MSTAR was released by the Defense Advanced Research Project Agency and the Air Force Research Laboratory. The data was collected by the Sandia National Laboratory STARLOS sensor platform. As a benchmark dataset evaluating SAT ATR performance, the MSATR contains a large number of SAR images including ten different types of ground targets, where tank, rocket launcher, armored personnel carrier, air defense unit, and bulldozer, are acquired as 1-ft resolution X-band SAR images in the range from $\text{0}{}^\circ$ to $\text{360}{}^\circ$. These targets are captured with different aspect angles, depression angles, and serial numbers. Ten different classes of ground targets used in our experiments are of different serial numbers with different depression angles. The SAR and corresponding optical images of ten types of targets are shown in Fig. \ref{sampleMSTAR}.
 
The OpenSARShip dataset aims to develop sophisticated ship detection and classification algorithms under high interference. The data was collected from 41 Sentinel-1 images under various environmental conditions. There are 17 types of SAR ships, 11346 ship chips in total, being integrated with automatic identification system (AIS) messages. The labels in this dataset are reliable because these labels of ships are based on AIS information \cite{OpenSARShip}. In our experiment, we adopt the ground range detected (GRD) data which are under Sentinel-1 IW mode with a resolution of $2.0m \times 1.5m$. For the size of ships, the length ranges from 92m to 399m and the width ranges from 6m to 65m. In the following experiments, both the VV and VH data are employed in the training, validation, and testing procedures. Fig. \ref{sampleOPEN} shows the sample SAR images of a three-class target.

The configurations of the training process and the network are presented here. The size of input SAR images is $224 \times 224$ by applying bilinear interpolation to the original data. The values of ${\lambda_1}$, ${\lambda_2}$ and ${\lambda_3}$ are set as 1, 0.5 and 0.5. The value of ${{\rm{margin}}}$ is set as 0.3. The values of ${v_1}$, ${v_2}$, ${v_3}$ and ${v_4}$ are set as 2, 2, 8 and 2 respectively. The proposed method is tested and evaluated on a GPU cluster with Intel(R) Xeon(R) CPU E5-2698 v4 @ 2.20GHz, eight Tesla V100 with eight 32GB memories. 
The proposed method is implemented using the open-source PyTorch framework with only one Tesla V100. 
The batch size is set as 128. The learning rate is initialized as 0.01 and reduced with the 0.5 ratios for every 25 epochs. There are also 10 epochs to warm up for training. Other hyperparameters are shown in Fig. \ref{fig3}. We choose swin-transformer as the backbone, which is more suitable for SAR image feature learning. Because SAR images are different from optical images, due to its specific imaging process, each scattering point of SAR images is related to the actual scene in a larger range. Therefore, the feature extraction models for SAR images more needs construction of long-distance information.

\subsection{Method Soundness Verification}

In this section, the soundness verification experiments of our method are evaluated. We run two experiments with 20 training samples each class under different network configurations. The first network configuration is the normal recognition network just without our proposed method, single upper yellow branch in Fig. \ref{fig3}. The second network configuration is our method shown in Fig. \ref{fig3}. The training and testing samples are from the MSTAR dataset, the distribution of the whole MSTAR dataset is shown in Table \ref{ttnumMSTAR}. The training data were collected at a $\text{17}{}^\circ$ depression angle and randomly sampled 20 each class, and the testing data were collected at a $\text{15}{}^\circ$ depression angle, the whole testing data are employed in the testing phase. We all trained the networks with two different configurations with 120 epochs and employed the parameters of the last epochs to shown the results.

\begin{figure*}
\centering
\includegraphics[width=0.85\textwidth]{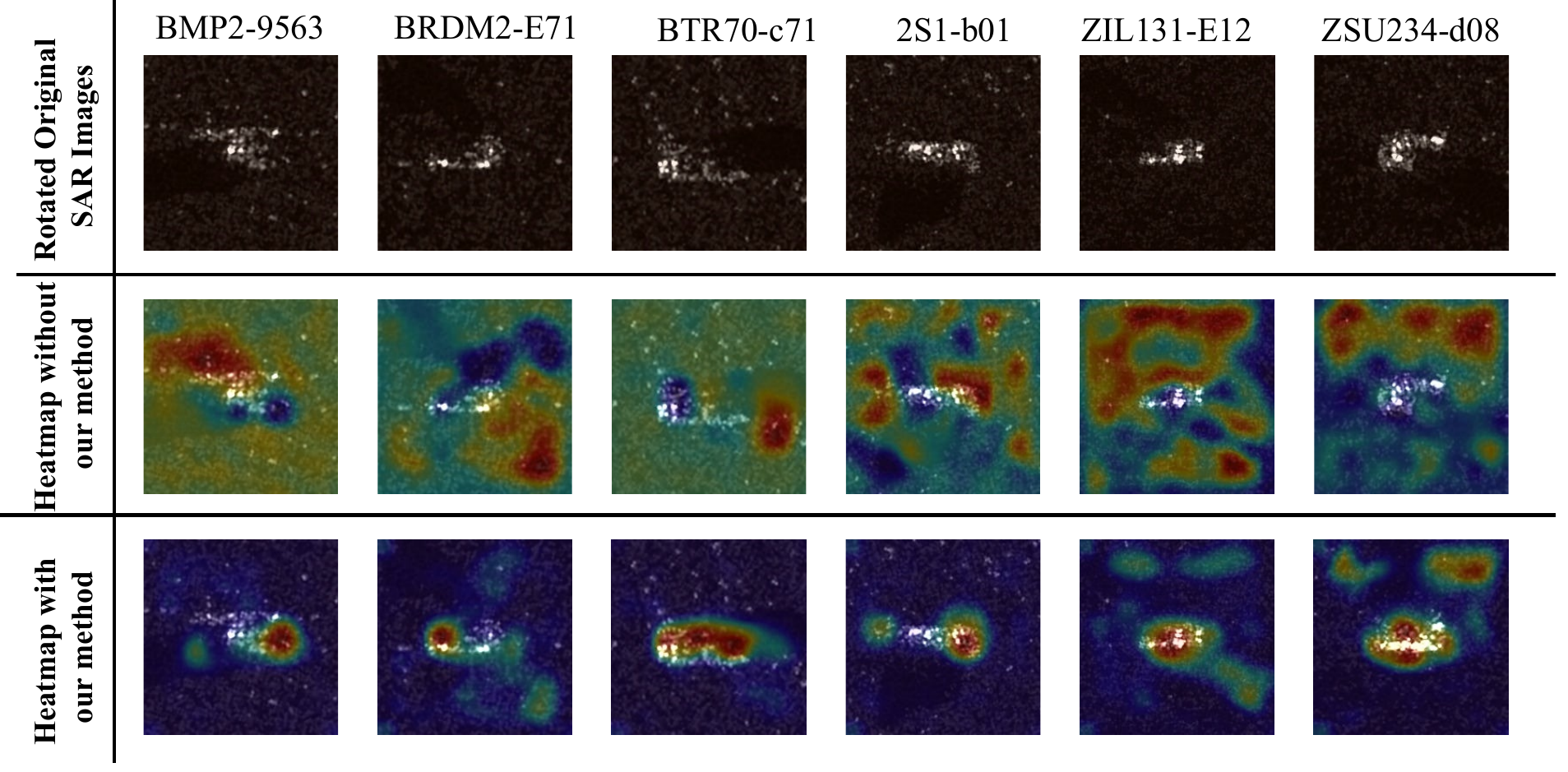}
\caption{Heatmaps of testing samples under two different network configurations with 20 training samples each class in MSTAR. The first network configuration is the normal recognition network just without our proposed method, single upper yellow branch in Fig. \ref{fig3}. The second network configuration is our method shown in Fig. \ref{fig3}. The recognition ratios without our method is 74.59\%, the recognition ratios of our methods is 94.97\%. Notion that in our method, the local image regions are generated by binarizing the heatmaps.}
\label{ablation}
\end{figure*}

First, the recognition ratios of the first configuration (without our method) is 74.59\%, the recognition ratios of our methods is 94.97\%.
Then, the heatmaps of two different configurations is shown in Fig. \ref{ablation}. It is clear that the first network configuration can't focus on the discriminative local image region, rather focus on the background which is useless for the recognition. Our method can capture the discriminative local image region in the whole image, discard other useless or even harmful image regions in the whole image.

From the comparison of two different network configurations in recognition ratios and heatmaps, 
the recognition ratios of our method obviously outperform the one of the network without our method. The heatmaps of two different network configurations also shows the superiority of our method and the reason of the higher recognition ratios.
Therefore, the soundness of our method has been validated.

\subsection{Recognition Results}
In this section, the recognition results of our method in MSTAR and OpenSARShip dataset are shown. The experiment configuration are described at first, then the results are presented and analysed.

\subsubsection{Recognition Results under SOC in MSTAR}

Under the MSTAR dataset including ten different targets, the recognition performance of our proposed SCDR is evaluated in the experimental setup of SOC. The training data were collected at a $\text{17}{}^\circ$ depression angle, and the testing data were collected at a $\text{15}{}^\circ$ depression angle. The distribution of training and testing images in the experimental setup are listed in Table \ref{ttnumMSTAR}. Note that in Table \ref{ttnumMSTAR}, the number of each target indicates the number of raw SAR images in the MSTAR dataset. In the following experiments, if given a 10-way X-shot experiment, the X here represents the number of randomly chosen images from the raw SAR images.

\begin{table}[]
\renewcommand{\arraystretch}{1.5}
\centering
\caption{Original Image Number of Different Depressions for SOC}
\label{ttnumMSTAR}
\begin{tabular}{c|cc|cc}
\hline\hline
\multirow{2}{*}{Class} & \multicolumn{2}{c|}{Training}            & \multicolumn{2}{c}{Testing}             \\ \cline{2-5} 
                       & \multicolumn{1}{c|}{Number} & Depression & \multicolumn{1}{c|}{Number} & Depression \\ \hline
BMP2-9563 & \multicolumn{1}{c|}{233} & \multirow{9}{*}{$\text{17}{}^\circ$} & \multicolumn{1}{c|}{195} & \multirow{9}{*}{$\text{15}{}^\circ$} \\ \cline{1-2} \cline{4-4}
BRDM2-E71              & \multicolumn{1}{c|}{298}    &            & \multicolumn{1}{c|}{274}    &            \\ \cline{1-2} \cline{4-4}
BTR60-7532             & \multicolumn{1}{c|}{256}    &            & \multicolumn{1}{c|}{195}    &            \\ \cline{1-2} \cline{4-4}
BTR70-c71              & \multicolumn{1}{c|}{233}    &            & \multicolumn{1}{c|}{196}    &            \\ \cline{1-2} \cline{4-4}
D7-92                  & \multicolumn{1}{c|}{299}    &            & \multicolumn{1}{c|}{274}    &            \\ \cline{1-2} \cline{4-4}
2S1-b01                & \multicolumn{1}{c|}{299}    &            & \multicolumn{1}{c|}{274}    &            \\ \cline{1-2} \cline{4-4}
T62-A51                & \multicolumn{1}{c|}{299}    &            & \multicolumn{1}{c|}{273}    &            \\ \cline{1-2} \cline{4-4}
T72-132                & \multicolumn{1}{c|}{232}    &            & \multicolumn{1}{c|}{196}    &            \\ \cline{1-2} \cline{4-4}
ZIL131-E12             & \multicolumn{1}{c|}{299}    &            & \multicolumn{1}{c|}{274}    &            \\ \cline{1-2} \cline{4-4}
ZSU234-d08             & \multicolumn{1}{c|}{299}    &            & \multicolumn{1}{c|}{274}    &            \\ \hline\hline
\end{tabular}
\end{table}

In Table \ref{soc}, the recognition performance of the proposed SCDR is demonstrated quantitatively. The first row lists the number of training images in each target class. The type of targets in the first column is given as the class and series linking with hyphens. It is noted that there is no additional training dataset or support dataset except k-shot for each class. The $N$ training images are augmented by $10$ times through randomly sampling $10$ image chips of $224 \times 224$ SAR images from a $384 \times 384$ SAR image after the bilinear interpolation, which ensures the central target is complete \cite{my1}.
All the recognition ratios of the ten targets in MSTAR and the average recognition ratio listed in Table \ref{soc} are calculated after 20 experiments.

\begin{table}[]
\renewcommand{\arraystretch}{1.5}
\setlength\tabcolsep{2.8pt}
\centering
\caption{Recognition Performance (\%) under SOC on MSTAR}
\label{soc}
\begin{tabular}{c|cccccccc}
\hline\hline
\multirow{2}{*}{Class} & \multicolumn{8}{c}{Labeled Number in Each Class}
\\ \cline{2-9}  & 1     & 2     & 5    & 10    & 20    & 25    & 40    & 80   \\ \hline
BMP2-9563       & 21.03 & 32.41 & 63.36 & 80.86 & 87.14 & 98.88 & 89.23 & 92.30 \\
BRDM2-E71       & 36.13 & 35.09 & 89.52 & 96.37 & 98.80 & 98.11 & 98.91 & 99.63 \\
BTR60-7532      & 46.67 & 60.53 & 69.14 & 95.26 & 95.36 & 96.48 & 94.87 & 95.89  \\
BTR70-c71       & 42.35 & 39.50 & 70.18 & 95.21 & 94.15 & 95.54 & 96.94 & 98.46 \\
D7-92           & 59.12 & 82.24 & 81.43 & 94.09 & 93.88 & 100.00 & 97.81 & 100.00 \\
2S1-b01         & 33.58 & 26.85 & 87.26 & 98.78 & 99.61 & 98.79 & 99.27 & 100.00 \\
T62-A51         & 37.36 & 40.24 & 81.60 & 92.00 & 93.88 & 96.06 & 99.27 & 98.90  \\
T72-132         & 37.24 & 57.24 & 60.40 & 84.07 & 94.00 & 95.79 & 99.49 & 99.48 \\ 
ZIL131-E12      & 41.97 & 50.95 & 80.36 & 97.63 & 95.04 & 94.52 & 99.27 & 100.00 \\
ZSU234-d08      & 74.09 & 65.59 & 70.28 & 67.08 & 97.78 & 94.44 & 100.00 & 100.00  \\ \hline
Average         & 43.75 & 48.74 & 74.64 & 88.58 & 94.97 & 96.74 & 97.81  & 98.72 \\ \hline\hline
\end{tabular}
\end{table}

It is clear that when the number of training samples is larger than 25 shots for each class, the recognition ratios can achieve more than 97.80\%. When the number of training sample are 80 and 40, the recognition ratios are 98.72\% and 97.81\%, respectively. When there are 25, 20 and 10 training samples, the average recognition ratios are 96.74\%, 94.97\% and 88.58\%, respectively.
When the training samples are limited to 5-shot and 2-shot, there are a total of 50 or 20 SAR images for training before augmentation. In this case, our proposed SCDR still obtains 74.64\% and 48.74\% recognition accuracy for ten-class recognition. The recognition ratios of BMP2-9563, T72-132 and T62-A51 are more largely influenced by the limited training samples than that of other class types.
From all the recognition ratios of the ten targets under 2-shot, part of them are still recognized. As the training samples are decreasing, the recognition performance of BMP2-9563, BTR60-7532, BRDM2-E7, T62-A51 and ZIL131-E12 are influenced more obviously than that of the other five target types. The proposed SCDR can obtain the uplifting average recognition ratios of 48.74\%, which is mainly decreased by the recognition of the six target types above.

From the experimental results and analysis above, the proposed SCDR can achieve outstanding recognition performance when the training samples are ranging from 2 to 80 for each target type in ten classes under SOC in MSTAR. 

\subsubsection{Recognition Results under EOCs in MSTAR}

The extended operating condition (EOC) consists of EOC- CV (configuration variant), EOC-D (depression variant) and EOC-VV (version variant). The experiments of EOCs are similar to the practical situation and hard to achieve high performance because of the large variance between the training samples and testing samples. 
When it comes to the limited training sample, the performance of EOCs is extremely hard to be improved.

It is known that variance in depression angle can dramatically exacerbate recognition performance. The testing dataset is captured at a $\text{30}{}^\circ$ depression angle and the training dataset is the SAR images of four corresponding classes at a $\text{17}{}^\circ$ depression angle as listed in Table \ref{ttnumEOCs}.
The summary of the testing data is listed in Table \ref{ttnumEOCs} and the recognition performance of EOC-D under limited training samples are listed in Table \ref{EOCDperf}. 

\begin{table}[tb]
\renewcommand{\arraystretch}{1.5}
\setlength\tabcolsep{4.2pt}
\caption{Training and Testing dataset under EOCs}
\centering
\label{ttnumEOCs}
\begin{tabular}{lcclcc}
\hline\hline
Train      & Number & \begin{tabular}[c]{@{}c@{}}Depression\\ Angle\end{tabular}  & Test(EOC-D) & Number & \begin{tabular}[c]{@{}c@{}}Depression\\ Angle\end{tabular} \\ \hline
2S1        & 299    & \multirow{4}{*}{$\text{17}{}^\circ$}    & 2S1-b01    & 288 & \multirow{4}{*}{$\text{30}{}^\circ$}   \\ 
BRDM2      & 298    &   & BRDM2-E71  & 287 &    \\ 
T72        & 232    &   & T72-A64    & 288 &    \\ 
ZSU234     & 299    &   & ZSU234-d08 & 288 &    \\ \hline\hline
Train      & Number & \begin{tabular}[c]{@{}c@{}}Depression\\ Angle\end{tabular}  & Test(EOC-C) & Number & \begin{tabular}[c]{@{}c@{}}Depression\\ Angle\end{tabular} \\ \hline
BMP2       & 233    & \multirow{5}{*}{$\text{17}{}^\circ$}    & T72-S7     & 419 & \multirow{5}{*}{$\text{15}{}^\circ$, $\text{17}{}^\circ$}    \\ 
BRDM2      & 298    &   & T72-A32    & 572 &    \\ 
BTR70      & 233    &   & T72-A62    & 573 &    \\ 
T72        & 232    &   & T72-A63    & 573 &    \\ 
           &        &   & T72-A64    & 573 &    \\ \hline\hline
Train      & Number & \begin{tabular}[c]{@{}c@{}}Depression\\ Angle\end{tabular}  & Test(EOC-V) & Number & \begin{tabular}[c]{@{}c@{}}Depression\\ Angle\end{tabular} \\ \hline
BMP2       & 233    & \multirow{7}{*}{$\text{17}{}^\circ$}    & T72-SN812  & 426 & \multirow{7}{*}{$\text{15}{}^\circ$, $\text{17}{}^\circ$}    \\ 
           &        &   & T72-A04    & 573 &    \\ 
BRDM2      & 298    &   & T72-A05    & 573 &    \\ 
           &        &   & T72-A07    & 573 &    \\ 
BTR70      & 233    &   & T72-A10    & 567 &    \\ 
           &        &   & BMP2-9566  & 428 &    \\ 
T72        & 232    &   & BMP2-C21   & 429 &    \\ \hline\hline
\end{tabular}
\end{table}

\begin{table}[]
\renewcommand{\arraystretch}{1.5}
\setlength\tabcolsep{7pt}
\centering
\caption{Recognition Performance (\%) under EOC-D on MSTAR}
\label{EOCDperf}
\begin{tabular}{c|cccccccc}
\hline\hline
\multirow{2}{*}{Class} & \multicolumn{6}{c}{Labeled Number in Each Class}
\\ \cline{2-7}  & 1     & 2     & 5     & 10    & 25    & 80        \\ \hline
BRDM2-E71       & 63.76 & 63.41 & 70.03 & 99.28 & 98.94 & 99.31     \\
2S1-b01         & 75.00 & 68.75 & 82.99 & 72.84 & 81.61 & 96.00     \\
T72-132         & 59.72 & 69.10 & 63.19 & 99.31 & 98.69 & 99.55     \\ 
ZSU234-d08      & 50.00 & 47.92 & 81.94 & 84.57 & 77.93 & 84.18     \\ \hline
Average         & 62.12 & 62.29 & 74.54 & 85.93 & 86.97 & 94.01     \\ \hline\hline
\end{tabular}
\end{table}

\begin{table}[]
\renewcommand{\arraystretch}{1.5}
\setlength\tabcolsep{7pt}
\centering
\caption{Recognition Performance (\%) under EOC-C on MSTAR}
\label{EOCCperf}
\begin{tabular}{c|cccccccc}
\hline\hline
\multirow{2}{*}{Class} & \multicolumn{6}{c}{Labeled Number in Each Class}
\\ \cline{2-7}  & 1     & 2     & 5     & 10    & 25    & 80    \\ \hline
T72-A32         & 58.74 & 63.99 & 84.09 & 81.12 & 81.12 & 87.24 \\
T72-A62         & 69.11 & 70.68 & 71.55 & 78.88 & 78.88 & 88.48 \\
T72-A63         & 67.36 & 71.38 & 69.11 & 69.98 & 69.98 & 81.33 \\
T72-A64         & 58.29 & 60.73 & 59.69 & 73.47 & 73.47 & 90.92 \\
T72-S7          & 59.67 & 60.38 & 90.21 & 94.27 & 94.27 & 100.00\\ \hline
Average         & 62.80 & 65.72 & 74.06 & 78.71 & 84.10 & 89.00 \\ \hline\hline
\end{tabular}
\end{table}

\begin{table}[tb]
\renewcommand{\arraystretch}{1.5}
\setlength\tabcolsep{7pt}
\centering
\caption{Recognition Performance (\%) under EOC-V on MSTAR}
\label{EOCVperf}
\begin{tabular}{c|cccccccc}
\hline\hline
\multirow{2}{*}{Class} & \multicolumn{6}{c}{Labeled Number in Each Class}
\\ \cline{2-7}  & 1     & 2     & 5     & 10    & 25    & 80    \\ \hline
BMP2-9566       & 22.43 & 29.21 & 51.40 & 62.38 & 80.61 & 96.50 \\
BMP2-C21        & 19.35 & 23.54 & 48.25 & 63.40 & 77.39 & 97.44 \\
T72-SN812       & 76.76 & 81.22 & 89.44 & 92.72 & 94.84 & 96.71 \\
T72-A04         & 68.41 & 61.95 & 78.01 & 76.44 & 83.25 & 88.66 \\
T72-A05         & 57.94 & 61.08 & 79.23 & 85.34 & 89.35 & 95.64 \\
T72-A07         & 65.45 & 63.35 & 74.69 & 82.20 & 91.80 & 96.16 \\
T72-A10         & 53.62 & 64.55 & 82.72 & 91.36 & 95.06 & 98.59 \\ \hline
Average         & 53.49 & 56.21 & 67.16 & 79.85 & 87.84 & 95.52 \\ \hline\hline
\end{tabular}
\end{table}

The average recognition ratios are more than 94.00\% when the training samples are equal to or larger than 80.
It means that the SCDR achieves a superior performance of recognition when the training samples are not particularly limited.
The average recognition ratios achieve 87.48\%, 86.96\% and 85.92\%, respectively, when the training samples are 40, 20 and 10.
When the training samples are decreasing from 80 to 10, the performance of SCDR is robust to the influence of the decreasing training samples.
When the training samples are limited to 5-shot and 2-shot, the recognition ratios can still achieve 74.54\% and 62.29\%. 

The recognition performance at the variance of target configuration and version (EOC-C and EOC-V) is also evaluated. 
The training data and the testing data for EOC-C and EOC-V are listed in Table \ref{ttnumEOCs}. 
Two different serial types of BMP2 and five different serial types of T72 captured at $\text{17}{}^\circ$ and $\text{15}{}^\circ$ depression angles are employed to evaluate the recognition performance under EOC-C. 
The recognition performance of EOC-C under limited training samples are listed in Table \ref{EOCCperf}. 
The average recognition ratios achieve above 84.00\% when the training samples are ranging from 25 to 80.
It is noted that when the training samples are limited to 10 for each class, the recognition ratios achieve 78.70\%. Furthermore, when the samples are decreasing to 5-shot, the SCDR gets the recognition ratio of 74.06\%, which means it is robust to decrease training samples in the range from 5 to 10.
When it is limited to 2-shot, the recognition ratio is 65.72\%. 
The results of EOC-C have illustrated the SCDR has the capability of handling the large variance of serial types with limited training samples.

There are four different serial type versions of T72 in the testing dataset being captured at $\text{17}{}^\circ$ and $\text{15}{}^\circ$ depression angles and utilized to evaluate the recognition performance under EOC-V. 
The recognition performance of EOC-V under limited training samples are listed in Table \ref{EOCVperf}. 
The average recognition ratios achieve above 95.50\% when the training samples are equal to or larger than 80 for each class. Even the training samples are greatly decreasing to 25, the recognition ratios still obtain 87.84\%.
The recognition ratios are close to 80.00\% when there are 10 images for each class. In other words, the SCDR classifies most of the testing samples correctly. 
When the training samples are limited in the range from 5 to 80, the recognition ratios decrease gradually. It is illustrated that the SCDR alleviate the problem of the limited training samples.

From the recognition performance of EOCs, it is clear that facing the larger variance of depression angels, configurations and type versions, the SCDR can still achieve good recognition performance. In other words, the SCDR are robust to the large variance between the imaging scenes of the training and testing samples, which increases the practical application capability of SAR ATR.

As shown by the recognition performance under SOC and EOCs in MSTAR data set, the proposed SCDR is robust and effective when confronting the large depression angle variant, configuration variant, and version variant, staying at a high-level recognition accuracy. 

\subsubsection{Recognition Results under OpenSARShip}

Following \cite{reduce1}, \cite{comparison4}, we choose three-class objects, bulk carrier, container ship, and tanks, from OpenSARShip. The distribution of training and testing data is listed in Table \ref{ttnumOPEN}. To avoid the effects of the imbalance of the number of three-class objects, concerning the least number of samples in all three classes with the training-testing ratio as $4:6$, we make the number of training samples the same for each class. The remaining samples are used for testing. In the experiments, the training images are adjusted to the size of $224 \times 224$ by bi-linear interpolation.

\begin{table}[]
\renewcommand{\arraystretch}{1.5}
\setlength\tabcolsep{15pt}
\centering
\caption{Image Number of Different Targets of OpenSARShip}
\label{ttnumOPEN}
\begin{tabular}{c|c|c|c}
\hline\hline
Class          & Training & Testing & Total  \\ \hline
Bulk Carrier   & 300      & 374     & 674  \\ 
Container Ship & 300      & 710     & 1010 \\ 
Tanks          & 300      & 253     & 553  \\ \hline\hline
\end{tabular}
\end{table}

\begin{table}[]
\renewcommand{\arraystretch}{1.5}
\centering
\caption{Recognition Performance (\%) under OpenSARship}
\label{performanceOPEN}
\begin{tabular}{c|cccccc}
\hline\hline
\multirow{2}{*}{Class} & \multicolumn{6}{c}{Labeled Number in Each Class}
\\ \cline{2-7} & 10    & 50    & 60    & 75    & 100   & 150   \\ \hline
Bulk Carrier   & 57.77 & 55.39 & 64.34 & 65.10 & 66.11 & 74.13 \\
Container Ship & 81.45 & 92.36 & 87.60 & 83.76 & 83.71 & 85.69 \\
Tanks          & 75.60 & 76.64 & 76.33 & 83.20 & 82.81 & 92.77 \\ \hline
Average        & 73.19 & 73.57 & 78.07 & 78.52 & 78.75 & 83.93 \\ \hline\hline
\end{tabular}
\end{table}

The recognition results of the proposed SCDR under OpenSARShip are presented in Table \ref{performanceOPEN}.
The recognition performance achieves above 78.75\% when training samples are sufficient, equal to or more than 100, 83.92\% under 150 for each target, and 78.75\% under 100 for each target.
From 10 to 100 training images for each class, the recognition ratios are ranging from 73.19\% to 78.75\%. More thoroughly, the recognition ratios are 78.75\% under 100 for each class, 78.52\% under 75 for each class, 78.06\% under 60 for each class, and 73.57\% under 50 for each class. 
Even when the training samples are limited to 10 training samples for each class, the recognition ratio still obtains 73.19\%. 
Therefore, the recognition performance is robust to the decreasing training samples from 100 to 10.

By analyzing the experimental results under the MSTAR and OpenSARShip datasets, our proposed SCDR shows the capabilities of robustness and effectiveness. It can be seen that when encountering cases of limited training samples and large variances in the experiments, the SCDR is capable of remaining good recognition performance even for different SAR datasets. To better and fairly illustrate the improvements in the recognition performance, the following subsection gives a quantitative comparison with other SAR ATR methods.

\begin{table}[]
\renewcommand{\arraystretch}{1.5}
\setlength\tabcolsep{10pt}
\centering
\caption{Comparison of Performance (\%) under SOC of MSTAR.}
\label{comparisonMSTAR}
\begin{tabular}{lcccc}
\hline\hline
\multirow{2}{*}{Algorithms} & \multicolumn{4}{c}{Image Number for Each Class}
\\ \cline{2-5} & 20    & 40    & 80    & All data  \\ \hline
PCA+SVM \cite{comparison1}               & 76.43             & 87.95             & 92.48             & 94.32                 \\
ADaboost \cite{comparison1}              & 75.68             & 86.45             & 91.45             & 93.51                 \\
LC-KSVD \cite{comparison1}               & 78.83             & 87.39             & 93.23             & 95.13                 \\
DGM \cite{comparison1}                   & 81.11             & 88.14             & 92.85             & 96.07                 \\
DNN1 \cite{comparison2}                  & 77.86             & 86.98             & 93.04             & 95.54                 \\
DNN2 \cite{comparison3}                  & 79.39             & 87.73             & 93.76             & 96.50                 \\
CNN1 \cite{comparison1}                  & 81.80             & 88.35             & 93.88             & 97.03                 \\
CNN2 \cite{comparison4}                  & 75.88             & -                 & -                 & -                     \\
CNN+matrix \cite{comparison4}            & 82.29             & -                 & -                 & -                     \\
GAN-CNN \cite{comparison1}               & 84.39             & 90.13             & 94.91             & 97.53                 \\
MGAN-CNN \cite{comparison1}              & 85.23             & 90.82             & 94.91             & 97.81                 \\
Semisupervised \cite{reduce1}   & 92.62             & 97.11             & 98.65             & -                     \\
Ours                                     & \textbf{94.97}    & \textbf{97.81}    & \textbf{98.72}    & -                     \\
\hline\hline
\end{tabular}
\end{table}

\subsection{Comparison}

\begin{table*}[htb]
\renewcommand{\arraystretch}{1.5}
\setlength\tabcolsep{10pt}
\caption{Comparison of Performance (\%) under SOC and EOCs of MSTAR.\\
(Other methods employed an additional dataset (SARSIM \cite{SARSIM}) in training besides MSTAR.)}
\centering
\label{comparisonMSTAR2}
\begin{tabular}{lccccc}
\hline\hline
\multicolumn{6}{c}{SOC} \\ \hline
Algorithms                          & 10-way 1-shot     & 10-way 2-shot     & 10-way 5-shot     & 10-way 10-shot    & 10-way 25-shot    \\ \hline
DeepEMD \cite{camp1}                & 36.19±0.46    & 43.49±0.44    & 53.14±0.40    & 59.64±0.39    & 59.71±0.31    \\
DeepEMD grid \cite{comp5}           & 35.89±0.43    & 41.15±0.41    & 52.24±0.37    & 56.04±0.31    & 57.89±0.24    \\
DeepEMD sample \cite{comp5}         & 35.47±0.44    & 42.39±0.42    & 50.34±0.39    & 52.36±0.28    & 55.02±0.22    \\
DN4 \cite{camp2}                    & 33.25±0.49    & 44.15±0.45    & 53.48±0.41    & 64.88±0.34    & 79.28±0.22    \\ 
Prototypical Network \cite{camp3}   & 40.94±0.47    & 54.54±0.44    & 69.42±0.39    & 78.01±0.29    & 84.96±0.22    \\
Relation Network \cite{comp6}       & 39.16±0.46    & 43.49±0.44    & 53.14±0.40    & 59.64±0.39    & 59.71±0.31    \\
DKTS-N  \cite{intro6}               & \textbf{49.26±0.48}    & \textbf{58.51±0.42}    & 72.32±0.32    & 84.59±0.24    & 96.15±0.08    \\ 
Ours                                & 43.75    & 48.74    & \textbf{74.64}    & \textbf{88.58}    & \textbf{96.74}    \\ \hline\hline
\multicolumn{6}{c}{EOC-D} \\ \hline
Algorithms                          & 4-way 1-shot      & 4-way 2-shot      & 4-way 5-shot      & 4-way 10-shot     & 4-way 25-shot     \\ \hline
DeepEMD \cite{camp1}                & 56.81±0.99    & 62.80±0.78    & 65.16±0.61    & 67.58±0.49    & 70.22±0.35    \\
DeepEMD grid \cite{comp5}           & 55.95±0.43    & 57.46±0.41    & 63.81±0.37    & 65.72±0.31    & 68.85±0.24    \\
DeepEMD sample \cite{comp5}         & 49.65±0.44    & 54.00±0.42    & 58.19±0.39    & 60.34±0.28    & 62.51±0.22   \\
DN4 \cite{camp2}                    & 46.59±0.83    & 51.41±0.69    & 58.11±0.49    & 62.15±0.43    & 65.14±0.37   \\ 
Prototypical Network \cite{camp3}   & 53.59±0.93    & 56.57±0.53    & 61.94±0.48    & 65.13±0.43    & 69.81±0.36    \\
Relation Netwrok \cite{comp6}       & 43.21±1.02    & 46.93±0.81    & 54.97±0.56    & 38.62±0.49    & 44.42±0.43   \\
DKTS-N \cite{intro6}                & 61.91±0.91    & \textbf{63.94±0.73}    & 67.43±0.48   & 71.09±0.41   & 78.94±0.31    \\ 
Ours                                & \textbf{62.11}    & 62.29    & \textbf{74.54}    & \textbf{85.93}    & \textbf{86.98}    \\ \hline
\multicolumn{6}{c}{EOC-C} \\ \hline
Algorithms                          & 4-way 1-shot      & 4-way 2-shot      & 4-way 5-shot      & 4-way 10-shot     & 4-way 25-shot     \\ \hline
DeepEMD \cite{camp1}                & 38.39±0.86    & 45.65±0.75    & 54.53±0.60    & 62.13±0.50    & 63.71±0.36   \\
DN4 \cite{camp2}                    & 46.13±0.69    & 51.21±0.62    & 58.14±0.54    & 63.08±0.51    & 69.66±0.46    \\ 
Prototypical Network \cite{camp3}   & 43.59±0.84    & 51.17±0.78    & 59.15±0.70    & 64.15±0.61   & 69.95±0.50    \\ 
Relation Netwrok \cite{comp6}       & 42.13±0.90    & 48.24±0.82    & 53.12±0.71    & 36.28±0.59    & 39.81±0.42    \\
DKTS-N \cite{intro6}                & 47.26±0.79    & 53.61±0.70   & 62.23±0.56    & 68.41±0.51    & 74.51±0.36    \\ 
Ours                                & \textbf{62.80}    & \textbf{65.72}    & \textbf{74.06}    & \textbf{78.71}    & \textbf{84.10}    \\ \hline
\multicolumn{6}{c}{EOC-V} \\ \hline
Algorithms                          & 4-way 1-shot      & 4-way 2-shot      & 4-way 5-shot      & 4-way 10-shot     & 4-way 25-shot     \\ \hline
DeepEMD \cite{camp1}                & 40.92±0.76    & 49.12±0.65    & 58.43±0.51    & 67.64±0.42    & 67.03±0.21    \\
DN4 \cite{camp2}                    & 47.00±0.72    & 52.21±0.61    & 58.87±0.55    & 63.93±0.52    & 70.64±0.47    \\ 
Prototypical Network \cite{camp3}   & 45.13±0.72    & 52.86±0.65    & 62.07±0.52    & 67.71±0.40    & 73.41±0.31    \\ 
Relation Netwrok \cite{comp6}       & 40.24±0.91    & 46.32±0.82    & 54.22±0.68    & 35.13±0.52    & 33.18±0.46    \\
DKTS-N \cite{intro6}                & 48.91±0.70    & 55.14±0.58    & 65.63±0.49    & 70.18±0.42    & 76.97±0.35    \\
Ours                                & \textbf{53.49}    & \textbf{56.21}    & \textbf{67.16}    & \textbf{79.85}    & \textbf{87.84}    \\ \hline\hline
\end{tabular}
\end{table*}

In this subsection, the performance of the SCDR are compared with other methods under two different ranges of training sample numbers. As described in related works, the methods for the limited SAR data only employ K-shot from MSTAR in training. If one method for the limited SAR data employs other images as unlabeled images or other usages, we will present more detail about the usages of the method. 

In Table \ref{comparisonMSTAR}, the comparison with state-of-the-art methods for the limited SAR data is presented. 
MGAN-CNN \cite{reduce1} improves the generated image quality of GAN by multi-discriminator architecture for the improvement of the recognition performance. CNN1 and GAN-CNN are the simplified versions of MGAN-CNN.
Semisupervised \cite{reduce1} proposed the self-consistent augmentation to utilize the unlabeled data. For a fair comparison, we used the results of recognition without sufficient unlabeled images in this paper.
The methods used in the above papers are also compared with the SCDR, including PCA+SVM, ADaboost, LC-KSVD, DGM, DNN-based methods (DNN1 and DNN2), and CNN-based methods (CNN2 and CNN+matrix) \cite{reduce1}. 

From the comparison, the recognition performance decreases obviously when the image number for each class is changed from all data to 20 samples. MGAN-CNN increased the performance lightly under 20 and 40 samples. Semisupervised boost the performance under 20, 40, and 80 samples with its self-consistent augmentation and training resources. From the quantitative comparison, it has illustrated that the SCDR obtain higher recognition performance than others under any number of training images in each class.

In Table \ref{comparisonMSTAR2}, the comparison with state-of-the-art methods from 1-shot to 25-shot is presented. 
DeepEMD \cite{camp1} employed the earth mover’s distance (EMD) as a metric to compute a structural distance between image representations to determine image relevance for classification. DeepEMD-Grid and DeepEMD-Sampling were novel structures with DeepEMD for slightly higher recognition performance.
Deep nearest neighbor neural network (DN4) \cite{camp2} employed a local descriptor-based image-to-class measure via a k-nearest neighbor searching over the deep local descriptors of convolutional feature maps. Prototypical network \cite{camp3} learns a metric space in which classification can be performed by computing distances to prototype representations of each class. By using both SARSIM \cite{SARSIM} and MSTAR, domain knowledge-powered two-stream deep network (DKTS-N) \cite{intro6} proposed a two-stream deep network for better recognition performance. 

It should be noted that the other methods in Table \ref{comparisonMSTAR2} employed an additional dataset (SARSIM) \cite{SARSIM} in training besides K-shot from MSTAR. 
From the comparison of SOC, the results of the SCDR are higher than other methods when the training samples for each class are larger than 2. Under the extreme conditions, like 1-shot and 2-shot, there are few training samples used in our method and no additional training datasets, which lead to the recognition ratios of the SCDR being relatively lower than the DKTS-N. When the training samples are increasing, like 5-shot or 10-shot, the effectiveness of our method is also becoming more apparent.
From the comparison of EOCs, the SCDR outperforms other methods under the recognition of the 4-way, even without additional training datasets. 
Clearly, the SCDR has a distinct improvement of 4-way 5-shot, 4-way 10-shot, and 4-way 25-shot under all three EOCs.
Under extreme conditions, the SCDR is still higher than other methods, only with 1-shot and 2-shot to train the model.

From the comparison with methods of the limited SAR data under SOC and EOCs in MSTAR, it has illustrated that the SCDR has achieved state-of-the-art performance among the methods of the limited SAR data, facing extreme few training samples and a large variance between the training and testing samples.

\begin{table}[]
\renewcommand{\arraystretch}{2}
\centering
\caption{Comparison of Performance (\%) under OpenSARShip}
\label{comparisonOPEN}
\begin{tabular}{c|ccc}
\hline\hline
\multirow{2}{*}{Methods} &
  \multicolumn{3}{c}{Number range of training images in each class} \\ \cline{2-4} 
 &
  \multicolumn{1}{c|}{1 to 50} &
  \multicolumn{1}{c|}{51 to 100} &
  101 to 240 \\ \hline
Semi-Supervised \cite{reduce1} &
  \multicolumn{1}{c|}{\begin{tabular}[c]{@{}c@{}}61.88 (20)\\ 64.73 (40)\end{tabular}} &
  \multicolumn{1}{c|}{68.67 (80)} &
  \begin{tabular}[c]{@{}c@{}}71.29 (120)\\ 74.96 (240)\end{tabular} \\ \hline
Supervised \cite{reduce1} &
  \multicolumn{1}{c|}{\begin{tabular}[c]{@{}c@{}}58.24 (20)\\ 62.09 (40)\end{tabular}} &
  \multicolumn{1}{c|}{65.63 (80)} &
  \begin{tabular}[c]{@{}c@{}}68.75 (120)\\ 70.83 (240)\end{tabular} \\ \hline
CNN\cite{comparison4} &
  \multicolumn{1}{c|}{62.75 (50)} &
  \multicolumn{1}{c|}{68.52 (100)} &
  73.68 (200) \\ \hline
CNN+Matrix\cite{comparison4} &
  \multicolumn{1}{c|}{72.86 (50)} &
  \multicolumn{1}{c|}{75.31 (100)} &
  77.22 (200) \\ \hline
Ours &
  \multicolumn{1}{c|}{\textbf{\begin{tabular}[c]{@{}c@{}}73.19 (10)\\ 73.57 (50)\end{tabular}}} &
  \multicolumn{1}{c|}{\textbf{\begin{tabular}[c]{@{}c@{}}78.07 (60)\\ 78.52 (75)\\ 78.75 (100)\end{tabular}}} &
  \textbf{83.93 (150)} \\ \hline\hline
\end{tabular}
\end{table}

The comparison with other state-of-the-art methods under OPENSARShip is also presented in Table \ref{comparisonOPEN}. In different number ranges of the training sample for each class, the SCDR outperforms other methods. When the training samples are 10 for each class, the performance of the SCDR is higher than the performance of others with 50 for each class. It has illustrated that the SCDR has increased obviously the recognition performance under OPENSARShip dataset.

From the comparisons of the two benchmark datasets, MSTAR and OPENSARShip, the effectiveness and practicality of the SCDR have been validated. Meanwhile, under different imaging conditions of different datasets, the SCDR can achieve state-of-the-art performance. Furthermore, under extremely few training samples and large variance between the training samples and testing samples, the SCDR is robust for various experimental conditions, which will boost the practical application of SAR ATR methods.

\section{Conclusion}
The limited training SAR images hinder the practical application of SAR ATR methods. Based on crucial feature capture and discrimination module, our SCDR is proposed to capture and enhance the golden key of image to improve the performance of SAR ATR under limited training data.
Through the novel framework and feature capture module, the SCDR automatically searches and captures the discriminative local image regions in the whole image while discarding other useless or even harmful image regions.
The discrimination module is proposed to enhance the overall and local features with more intra-class compactness and inter-class separateness.
Finally, the overall and local features cooperated with learnable voting weights to finish the final recognition. 
The experimental results and comparisons on MSTAR and OPENSAR show that our method has achieved the best recognition performance to date. It has also illustrated that the SCDR is robust to the different large variances between the training samples and testing samples. 

\bibliographystyle{IEEEtran}
\bibliography{references}

\newpage

\begin{IEEEbiography}[{\includegraphics[width=1in,height=1.25in,clip,keepaspectratio]{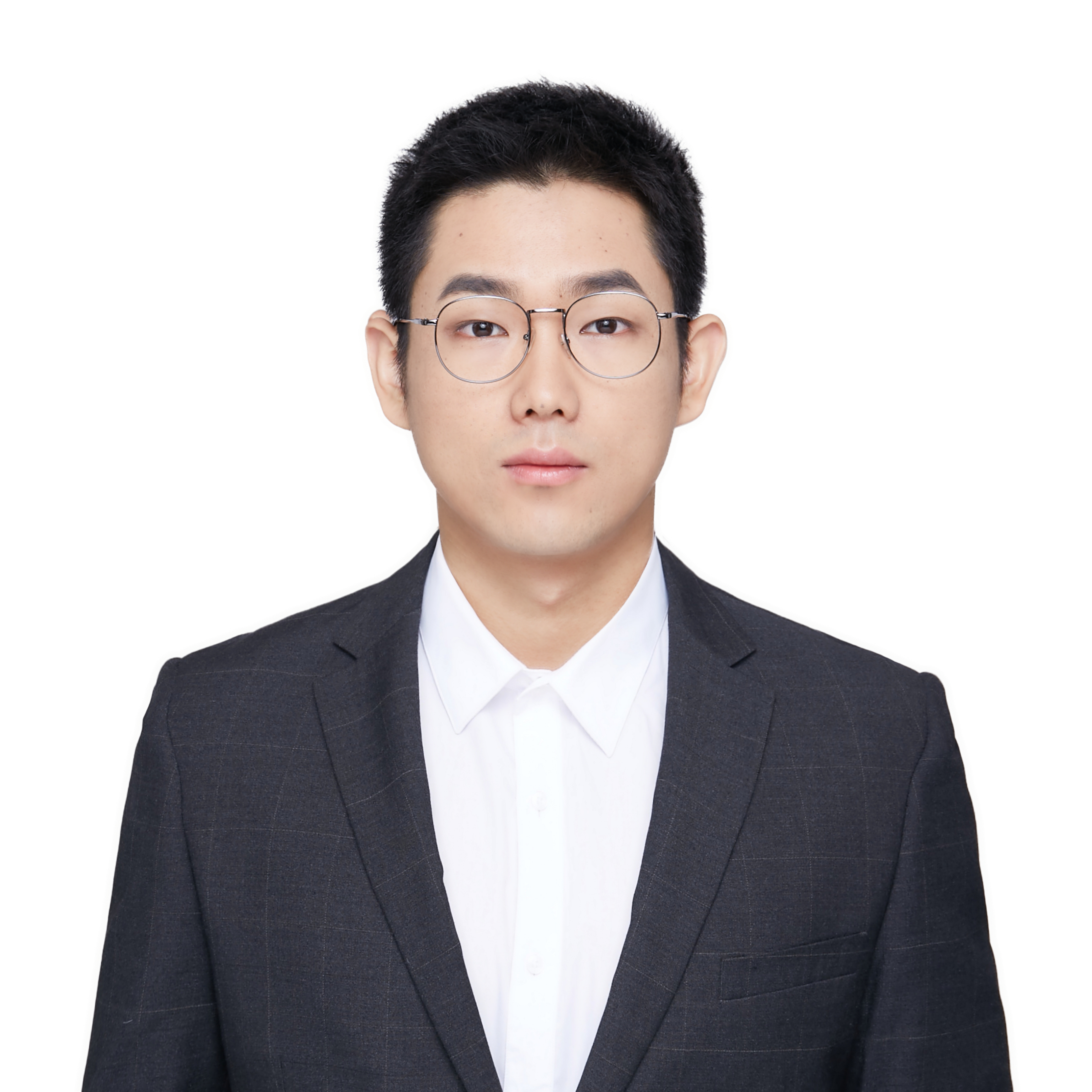}}]{Chenwei Wang}
received the B.S. degree from the School of Electronic Engineering, University of Electronic Science and Technology of China (UESTC), Chengdu, China, in 2018. He is currently pursuing the Ph.D. degree with the School of Information and Communication Engineering, University of Electronic Science and Technology of China, Chengdu, China.
His research interests include radar signal processing, machine learning, and automatic target recognition.
\end{IEEEbiography}

\begin{IEEEbiography}[{\includegraphics[width=1in,height=1.25in,clip,keepaspectratio]{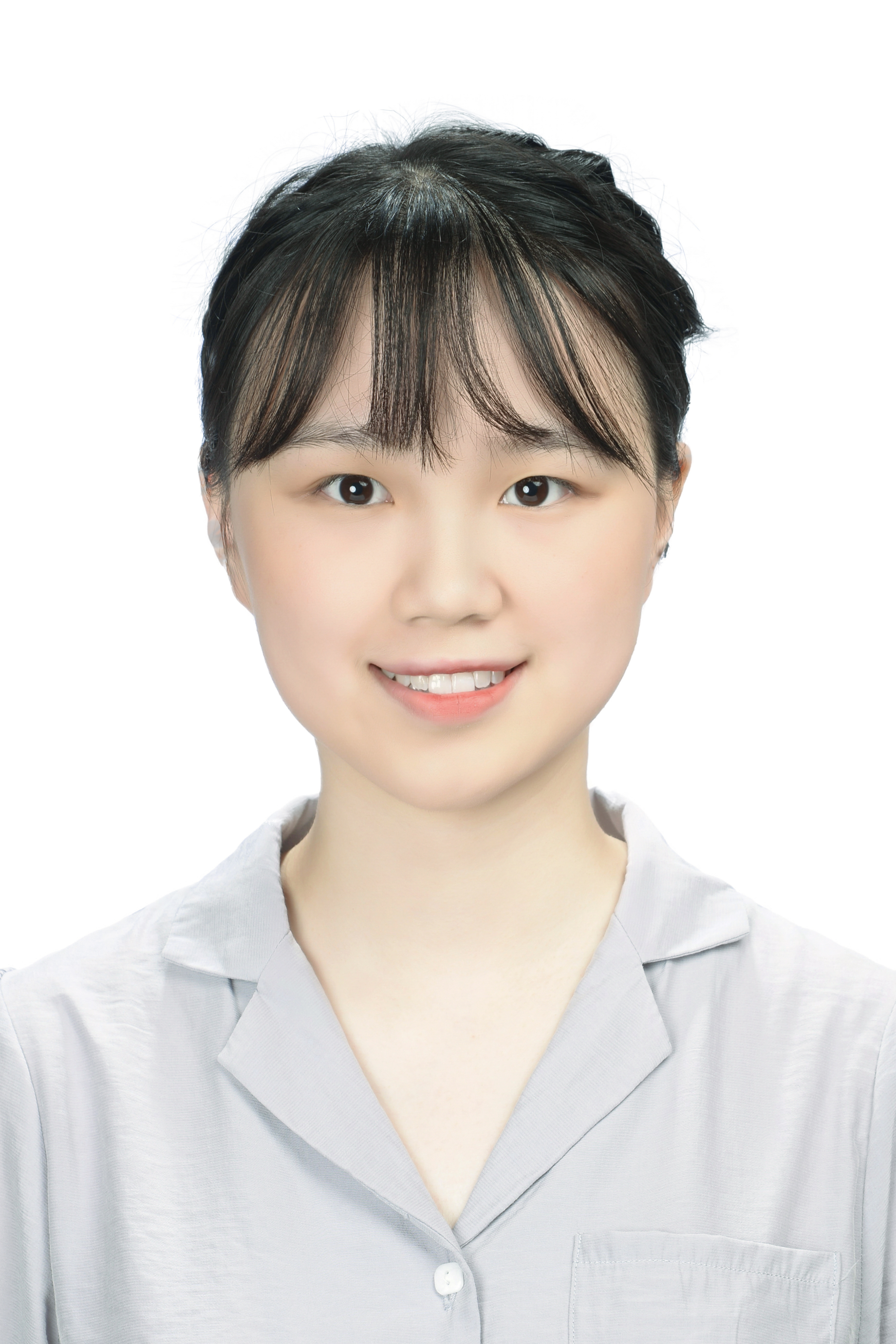}}]{Siyi Luo}
received the B.S. dual degree from Chongqing University (CQU), Chongqing, China, in 2021. She is currently pursuing the M.S. degree with the School of Information and Communication Engineering, University of Electronic Science and Technology of China, Chengdu, China. Her research interests include machine learning, target detection, and automatic target recognition.
\end{IEEEbiography}

\begin{IEEEbiography}[{\includegraphics[width=1in,height=1.25in,clip,keepaspectratio]{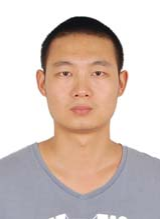}}]{Jifang Pei}
(M'19) received the B.S. degree from the College of Information Engineering, Xiangtan University, Hunan, China, in 2010, and the M.S. degree from the School of Electronic Engineering, University of Electronic Science and Technology of China (UESTC), Chengdu, China, in 2013. He received the Ph.D. degree from the School of Information and Communication Engineering, UESTC, in 2018. From 2016 to 2017, he was a joint Ph.D. Student with the Department of Electrical and Computer Engineering, National University of Singapore, Singapore. He is currently an Associate Research Fellow with the School of Information and Communication Engineering, UESTC. His research interests include radar signal processing, machine learning, and automatic target recognition.
\end{IEEEbiography}

\begin{IEEEbiography}[{\includegraphics[width=1in,height=1.25in,clip,keepaspectratio]{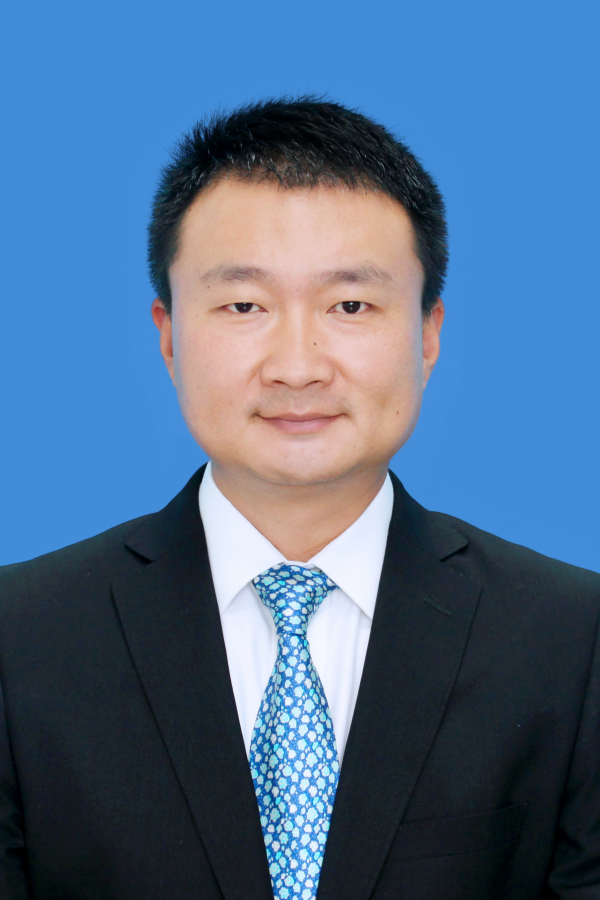}}]{Yulin Huang}
(M'08) received the B.S. and Ph.D. degrees from the School of Electronic Engineering, University of Electronic Science and Technology of China (UESTC), Chengdu, China, in 2002 and 2008, respectively. He is currently a Professor at the UESTC. His research interests include radar signal processing and SAR automatic target recognition.
\end{IEEEbiography}

\begin{IEEEbiography}[{\includegraphics[width=1in,height=1.25in,clip,keepaspectratio]{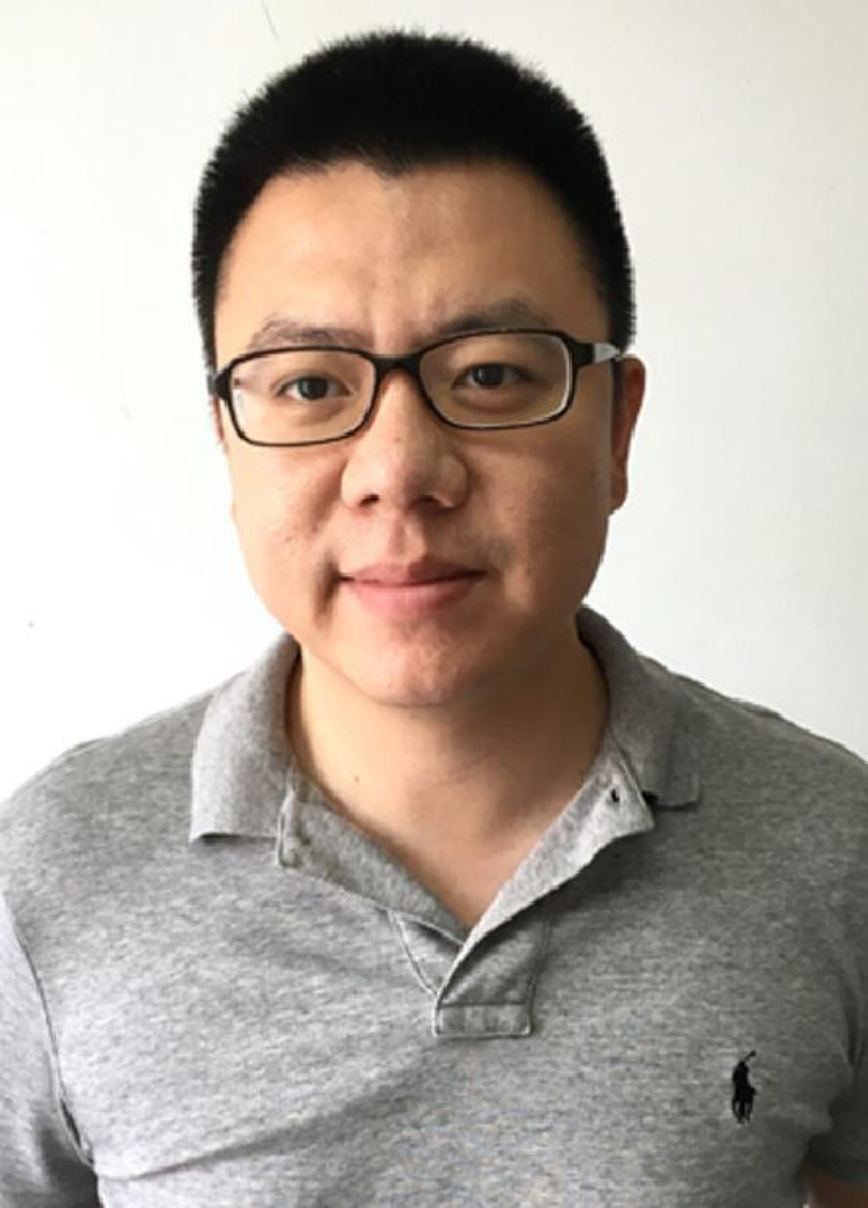}}]{Yin Zhang}
(M'16) received the B.S. and Ph.D. degrees from the School of Electronic Engineering, University of Electronic Science and Technology of China (UESTC), Chengdu, China, in 2008 and 2016, respectively. From September 2014 to September 2015, he had been a Visiting Student with the Department of Electrical and Computer Engineering, University of Delaware, Newark, USA. He is currently an Associate Research Fellow at the UESTC. His research interests include signal processing and radar imaging.
\end{IEEEbiography}

\begin{IEEEbiography}[{\includegraphics[width=1in,height=1.25in,clip,keepaspectratio]{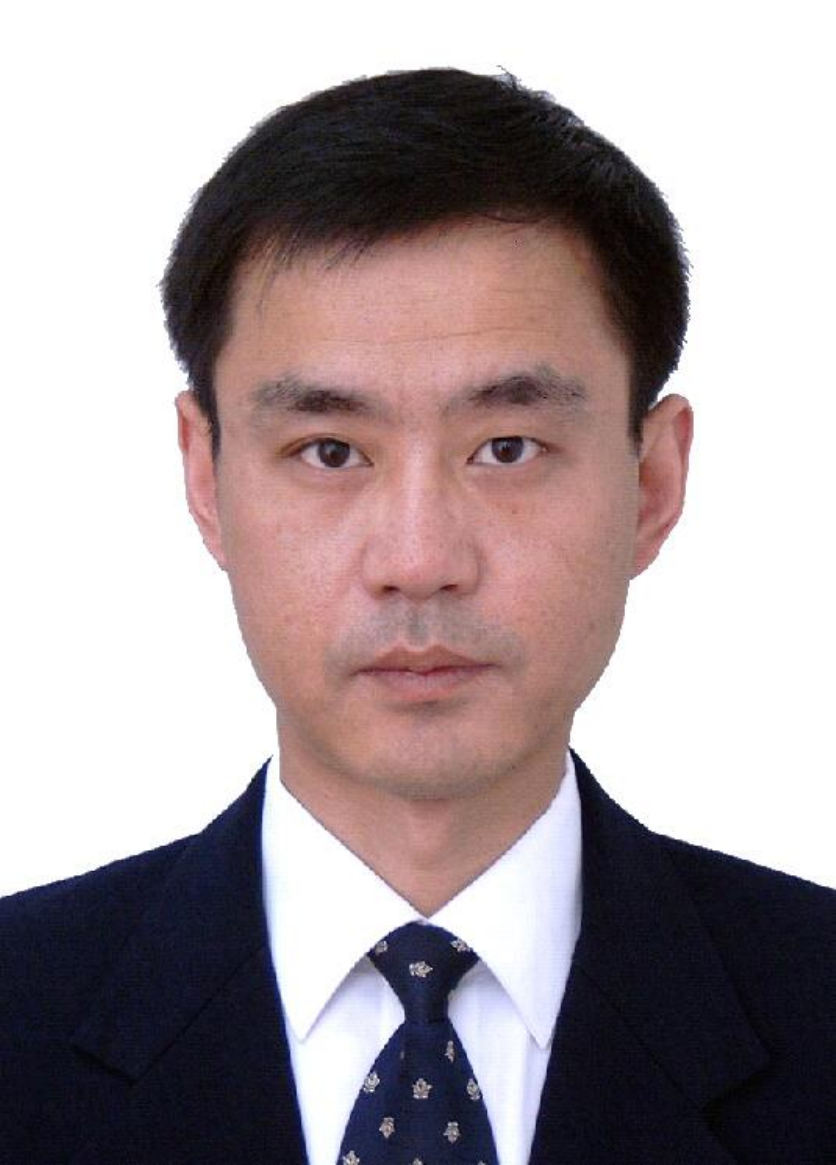}}]{Jianyu Yang}
(M'06) received the B.S. degree from the National University of Defense Technology, Changsha, China, in 1984, and the M.S. and Ph.D. degrees from the University of Electronic Science and Technology of China (UESTC), Chengdu, China, in 1987 and 1991, respectively. He is currently a Professor at the UESTC. His research interests include synthetic aperture radar and statistical signal processing. Prof. Yang serves as a Senior Editor for the Chinese Journal of Radio Science and the Journal of Systems Engineering and Electronics.
\end{IEEEbiography}

\vfill

\end{document}